\documentclass[%
 reprint,
 amsmath,amssymb,
 aps,
 showkeys,
]{revtex4-2}

\usepackage{graphicx}
\usepackage{amsmath}

\begin{document}

\preprint{APS/123-QED}

\title{Chaos in QCD? Gap equations and their fractal properties.}

\author{Thomas Kl\"ahn}
 \email{thomas.klaehn@csulb.edu}
 \affiliation{Department of Physics $\&$ Astronomy,
California State University Long Beach, Long Beach, CA
  90840, U.S.A. }
\author{Lee C. Loveridge}%
 \email{loverilc@piercecollege.edu}
 \affiliation{Los Angeles Pierce College, Woodland Hills, CA 91371, U.S.A.}
\author{Mateusz Cierniak}%
 \affiliation{Institute of Theoretical Physics, University of Wroc{\l}aw, 50-204 Wroc{\l}aw, Poland}

\date{\today}

\begin{abstract}
We discuss how iterative solutions of QCD inspired gap-equations at finite chemical potential show domains of chaotic behavior as well as non-chaotic domains which represent one or the other of the only two - usually distinct - positive mass gap solutions with broken or restored chiral symmetry, respectively. 
In the iterative approach gap solutions exist which exhibit restored chiral symmetry beyond a certain dynamical cut-off energy. 
A chirally broken, non-chaotic domain with no emergent mass poles and hence with no quasi-particle excitations exists below this energy cut-off.
The transition domain between these two energy separated domains is chaotic.
As a result, the dispersion relation is that of quarks with restored chiral symmetry, cut at a dynamical energy scale, determined by fractal structures. 
We argue that the chaotic origin of the infrared cut-off could hint at a chaotic nature of confinement and the deconfinement phase transition.
\end{abstract}

\maketitle

\section{Introduction}
In the early 1980's Benoit Mandelbrot pioneered the methodical study and computational visualization of the iteration of quadratic functions and began to cartograph the emerging fractal landscape \cite{doi:10.1111/j.1749-6632.1980.tb29690.x} which subsequently has been named in his honor as the Mandelbrot set.
With the advance of personal computers during the mid 80's fractals gained broad attention, scientifically as well as in popular science.

In 1986, Leo Kadanoff, in an article with the title "Fractals: Where's the physics?"\cite{Kadanoff:1986}, expressed concerned curiosity about an understanding of fractal properties in physics which goes beyond the identification of fractal dimensions for certain problems. Kadanoff stated that without a better understanding of how physical mechanisms result in geometrical form it is difficult to trace {\it types of questions with interesting answers.}
We wish to add that even in lack of such a deep understanding it is of course possible to find this kind of questions, as mentioned by Mandelbrot: {\it "I was asking questions which nobody else had asked before, because nobody else had actually looked at certain structures." }\cite{WoS}.

An example for this explorative approach is Hofstadter's butterfly which is  publicly less known. In 1976, ten years before Kadanoff asked his curious question and four years before Mandelbrot's famous work on the quadratic map, Douglas Hofstadter observed what he called {\it recursive structure} in the computed spectrum of electrons in electromagnetic fields \cite{PhysRevB.14.2239} which has been named after the visual appearance as {\it Hofstadter's butterfly}. A first experimental confirmation of this theoretical prediction has been reported nearly twenty years later in 1997 \cite{PhysRevLett.80.3232}.

There is no strict definition of what a fractal is but still most people would know one when they see it. Common descriptors of fractals refer to their non-analycity, self-similarity, non-linearity, iterative origin, chaotic behavior, non-integer (Hausdorff and other) dimension, to name a few. This paper has been motivated by the fact that QCD's gap equations are by definition highly non-linear and self-consistent. Self-consistency equates the quantity of interest, our gaps, to a functional which depends on these gaps themselves.
QCD's gap equations are organized in a hierarchy of inter-dependencies of an infinite number of n-point Green-functions and it is at the heart of contemporary approaches in this field to identify methods which reduce this infinite number in a manageable way while preserving key features of QCD like dynamical mass generation and confinement.
While one can argue how to obtain physically meaningful gap equations, viz. which set of approximations, truncations, etc. is the most reasonable, the self-consistent nature of these equations is not debated.
Already at the seemingly simple level of 2-point Green functions for a single quark flavor
appropriate truncation schemes allow to compute the mass spectrum of confined and deconfined quarks. The same methods allow for computation of meson and baryon spectra. Nothing of this is new and, although neither trivial nor brought to a final solution, it is in a structural sense reasonably well understood and dealt with in Dyson and Schwinger's  functional approach which proved to be a powerful tool to investigate the theories of QCD and QED. We refer to recent reviews for examples and more detailed information \cite{Roberts:2015lja,Horn:2016rip,Eichmann:2016yit,Burkert:2017djo,Fischer:2018sdj,Roberts:2020udq,Qin:2020rad,Barabanov:2020jvn}.

Practitioners in the field of Dyson-Schwinger equations frequently deal with problems that can arise from their self-consistent nature. As an example, one technique to solve gap equations is by means of iteration starting from an initial guess. There is no guarantee for the convergence of such an iteration in general nor that the obtained solution is physical. In order to cover 'all possible' solutions in this approach one would scan over different initial guesses.
Typically, one can 'tame' diverging iterations by damping the impact of the iteration itself. Instead of $g=F[g]$ one can write $g=\alpha g + (1-\alpha)F[g]$ where $g$ is the gap, $F$ a functional of the gap, and $\alpha$ a damping parameter close to but less than one, thus avoiding strong responses of $g$ to  the iteration. 
One can wonder - we claim one should - whether it is justified to apply such an algorithm.
It looks innocent in that sense that technically any solution of the original gap equation is a solution of the damped iteration equation. Nevertheless, at identical initial value both may give different answers and thus one can claim that the damping parameter might bear unwanted physical significance as it has been introduced ad hoc. What happens if the gap equation is allowed to iterate itself freely? We found only one, recently published, paper which asks exactly this question and comes to a clear conclusion: if the system is strongly coupled, chaos emerges and one can observe an infinite spectrum of 'unexpected' gap solutions with increasing coupling strength\cite{Martinez:2019ift}. 
In the paper we present, we give a brief explanation why these unexpected solutions actually should be expected.
Further, we employ a model with momentum dependent gap solutions. In an iterative and inherently fractal context this led us on a surprising journey which answered not all but plenty of the questions we asked and at the end of which we are left to wonder whether looking at QCD as a fractal theory might be a key to understand confinement as an emergent fractal phenomenon.

Section \ref{SEC:roots} briefly motivates how iterative mapping generates new solutions of an equation while preserving the solutions of the non-iterated 'seed' equation, Section \ref{SEC:MN} reviews the quark matter model by Munczek and Nemirovski (MN) in an extension for dense quark matter. We chose it for our exploration as it exhibits confinement and dynamical chiral symmetry breaking while being sufficiently simple to make it well suited for iterative mapping and analytic treatment. The following Section \ref{SEC:CHAOS} illustrates and cartographs chaotic features which emerge upon iteration of the gap equation, Section \ref{SEC:POLES} is a cautious attempt to interpret physical meaning into the interplay of chaotic and non-chaotic structures we observe. Our study focuses at the structure of the mass pole. Appearance and disappearance of the mass pole are highly driven by chaotic behavior. Further, the mass gap itself is allowed to switch between different, usually distinct solutions. To our surprise, the physical properties of the iterative solutions provide a reasonable picture of how de-confinement could present itself in a model which possesses a gap equation with a single solution only.
Finally, we estimate how a finite width gluon interaction could affect the observed behavior of the quark dispersion relation under iteration in Sec.\ref{SEC:WIDTH} before we conclude in Sec.\ref{SEC:CONCL}.

\section{Self-consistency and the emergence of new roots amongst the old}
\label{SEC:roots}
We investigate possible consequences of chaos that appears in iterative solutions of non-linear and self-consistent equations in the complex domain.
For clarity of what we consider physics and what is math we start with the latter and briefly review Mandelbrot's fractal which is obtained by the iteration $z\xleftarrow{z_0=0;n\to\infty}f(z)$ with the explicit choice $f(z) = z^2+c$ to obtain the Mandelbrot set.
We chose to use the symbol $\xleftarrow{z_0;n} $ to have a distinguished notation for the iterative mapping process - specifying the number of iterations $n$ and the initial value $z_0$ - over the equal sign $=$ which appears in the analytic equation $z = z^2+c$.
It is worthwile to look at the differences between these two.
First, the polynomial equation has exactly two solutions $z_{1,2}$ for any given $c$ which are defined by the roots of the polynomial $P(z)=f(z)-z=z^2+c-z$. It is further easily seen that one can determine $c$ for a desired root $z_0$. 
For example, $P(z_0=0)=0$ if $c=0$.

In the iterative approach, each iteration generates a new polynomial,
\begin{eqnarray*}
    f_1(z,c)=\left(z\xleftarrow{z,1}z^2+c\right)&=&f(z)=z^2+c,\nonumber\\
    f_2(z,c)=\left(z\xleftarrow{z,2}z^2+c\right)&=&f(f(z))=\left(z^2+c\right)^2+c,\nonumber\\
    ...\nonumber\\
    f_m(z,c)=\left(z\xleftarrow{z,m}z^2+c\right)&=&f(f(..f(z)))\\
    &=&f_{m-1}^2(z,c)+c,\nonumber\\
\end{eqnarray*}
etc. ad infinitum. There is one trivial but fascinating property of this infinite set of equations which essentially inspired the presented work.
The left hand side of each of the previous equations has been set to $f_i(z,c)=z$ in order to obtain the next iteration  $f_{i+1}(z,c)=z$.
It is thus safe to state that the roots of $P_1(z,c)=f_1(z,c)-z$ are guaranteed to be roots of $P_2(z,c)=f_2(z,c)-z=f(f_1(z,c))+c$. As $P_2(z,c)$ is a 4th order polynomial, there are two more roots which, of course, did not appear for $P_1(z,c)$, a 2nd order polynomial. The important lesson to be learned is that for a self consistent non-linear equation $z=f(z)$ the iteration $z\xleftarrow{z;n} f(z)$ generates a new self consistent equation.
{\it While the solutions of the non-iterated equation remain a subset of solutions of the iterated equation,  the iterated equations can develop additional solutions.}

This is a peculiar, almost awkward situation, if one wishes to assign physical meaning to the original solutions of the equation $f(z)=z$. What makes these roots superior with respect to any of the iterative clones if all, original and clones, {\it share} these very same original solutions? Evidently, there is an infinite number of (iterated) functions which share the original roots. 
Is the original function with {\it only} these roots a superior or inferior function? Is it worth to ponder the meaning of the additional roots of iterated clones? Can we {\it safely} omit them? Do we miss important information when we ignore the duality of the gap equation as root defining equation and mapping rule?
We decided to explore and ponder possible meaning.

\section{The Munczek-Nemirovsky model}
\label{SEC:MN}

One approach to move towards an understanding of QCD is based on evaluating QCD's partition function by testing its response to external sources. This is the Dyson-Schwinger formalism which results in sets of coupled n-point Green functions. 
Out of these we are interested in the quark propagator which is obtained from the gap eguation
\begin{equation}
    S(p;\mu)^{-1}=i\vec\gamma\cdot\vec p+i\gamma_4(p_4+i\mu)+m+\Sigma(p;\mu),
\end{equation}
with the self-energy
\begin{eqnarray}
    \Sigma(p;\mu)&=&\int\frac{{\rm d}^4q}{(2\pi)^4}g^2(\mu)
    D_{\rho\sigma}(p-q;\mu)\nonumber\\
    &&\times\frac{\lambda^a}{2}\gamma_\rho S(q;\mu)
    \Gamma_\sigma^a(q,p;\mu).
\end{eqnarray}
Here, $m$ is the quark bare mass, $D_{\rho\sigma}(p-q;\mu)$ is the dressed gluon propagator and $\Gamma_\sigma^a(q,p;\mu)$ is
the dressed quark-gluon vertex.
This is the first of an infinite tower of gap equations which, without further approximations, couple back to this one. Further, there are similar equations for the dressed gluon-propagator and the quark-gluon vertex.
Note that the gap equation is a self-consistent non-linear (in most cases integral) equation, $S^{-1}=F[S]$.

Within the Munczek-Nemirovsky model \cite{PhysRevD.28.181} the dressed quark-gluon vertex is approximated by the free quark-gluon vertex, $\Gamma_\sigma^a(q,p;\mu)=\frac{\lambda^a}{2}\gamma_\sigma$. Gap equations applying this approximation are referred to as rainbow gap equations. For the dressed gluon propagator the model is specified by the choice
\begin{equation}
    g^2(\mu)D_{\rho\sigma}(k;\mu)=
    \left(
        \delta_{\rho\sigma}-\frac{k_\rho k_\sigma}{k^2}
    \right)
    4\pi^4\eta^2\delta^4(k).
\end{equation}
Due to the $\delta$-function which in configuration space corresponds to a constant, this is a very simplified approximation of the  gluon-propagator, specified by the coupling strength we set to $\eta=1.09$ GeV in accordance with \cite{PhysRevD.28.181}.
For non-zero relative momentum $k$ the interaction strength in this model vanishes, thus making it super-asymptotically free. Furthermore, the infrared enhanced $\delta$-function is sufficient to provide for dynamical chiral symmetry breaking and confinement, both features of QCD which we wish to address.
Finally, the $\delta$-function effectively turns the integral gap equation into an algebraic equation which can be solved analytically.

In order to obtain these solutions for the in-medium dressed-quark propagator one employs the general solution
\begin{eqnarray}
       S(p;\mu)^{-1}=
       i\vec\gamma\cdot\vec p A(\vec p^2,p_4)\nonumber\\
       +i\gamma_4(p_4+i\mu) C(\vec p^2,p_4)
       +B(\vec p^2,p_4).
\end{eqnarray}
Substitution into the dressed-quark gap-equation and appropriate tracing over the Dirac $\gamma$-matrices results in three coupled gap equations of which two (for $A$ and $C$) are identical:
\begin{eqnarray}
A(p,\mu)&=&1+\frac {\eta^2} 2 \frac {A(p,\mu)} {\tilde{p}^2 A^2(p,\mu)+B^2(p,\mu)}
\label{Agap}
\\
B(p,\mu)&=&m+ {\eta^2} \frac {B(p,\mu)} {\tilde{p}^2 A^2(p,\mu)+B^2(p,\mu)}.
\label{Bgap}
\end{eqnarray}
We introduced $\tilde p^2=\vec p^2+(p_4+i\mu)^2$.
In the chiral limit ($m=0$) one finds two distinct sets of solutions, one of them chirally symmetric and referred to as the Nambu phase,
\begin{eqnarray}
A(p,\mu)&=&\frac 1 2 \left( 1 \pm \sqrt{1+\frac {2\eta^2} {\tilde{p}^2}}\right) \\
B(p.\mu)&=&0
\end{eqnarray}
whereas for the other solution, the Wigner phase, chiral symmetry is broken for $\mathcal{R}(\tilde p^2)<\eta^2/4$,
\begin{eqnarray}
A(p,\mu)&=&2 \\
B(p,\mu)&=&\sqrt{\eta^2-4 \tilde{p}^2}.
\end{eqnarray}
For $\mathcal{R}(\tilde p^2)>\eta^2/4$ the gap solution of the Wigner phase coincides with the Nambu solution.
Note, that these solutions are obtained in Euclidean metric, 
but hold in Minkowski metric after a simple transformation, $\tilde p^2_E \to \tilde p^2_M$, with
$\tilde p^2_E = \vec p^2+(p_4+i\mu)^2$ and $\tilde p^2_M=\vec p^2-(p_4+i\mu)^2$.
Due to our interest in particle mass poles
our investigation of the model is performed in Minkowski metric.
For the next section however, the specific metric is not relevant; we will only work with the fact that $\tilde p^2$ is complex-valued and thus can be decomposed into real and imaginary part, viz. $\tilde p^2=z^2_R+i z^2_I$. We chose to label real and imaginary part with squared quantities as a reminder that they come in units of energy square.

\section{Iterative Chaos}
\label{SEC:CHAOS}

Gap equations (\ref{Agap}) and (\ref{Bgap}) lead to  4$^{th}$ order polynomial equations with up to four distinct complex valued solutions at given $\tilde p^2$ for each gap. 
Generally, this is the whole solution space one would consider and the only task left is to identify the one physical solution.
However, the self-consistent nature of (\ref{Agap}) and (\ref{Bgap}) is evident and, according to our reasoning in the previous section there is a possibility for iterated functions with the same four and additional solutions.

Before we discuss our analysis, a few comments should be made. Defined by contact interaction in momentum space we chose a very simple model for the effective gluon propogator. For dressed-gluon propagators with finite width in momentum space the corresponding gap equations turn into integral equations.
Thus, momenta couple and the simplicity of the MN model which we take advantage of for this  exploration is  lost. We address this issue in more detail in the last section of this paper.

As sketched in the Introduction, for models with sophisticated non-trivial interaction-kernel, iteration is a practical path to find gap solutions.
We outlined before, that this leaves us with the possibility that iteration generates new functions which possess roots that correspond to solutions of the original gap equation and potentially an infinite number of additional roots.

\begin{figure}[htb]
\includegraphics[width=.5\textwidth]{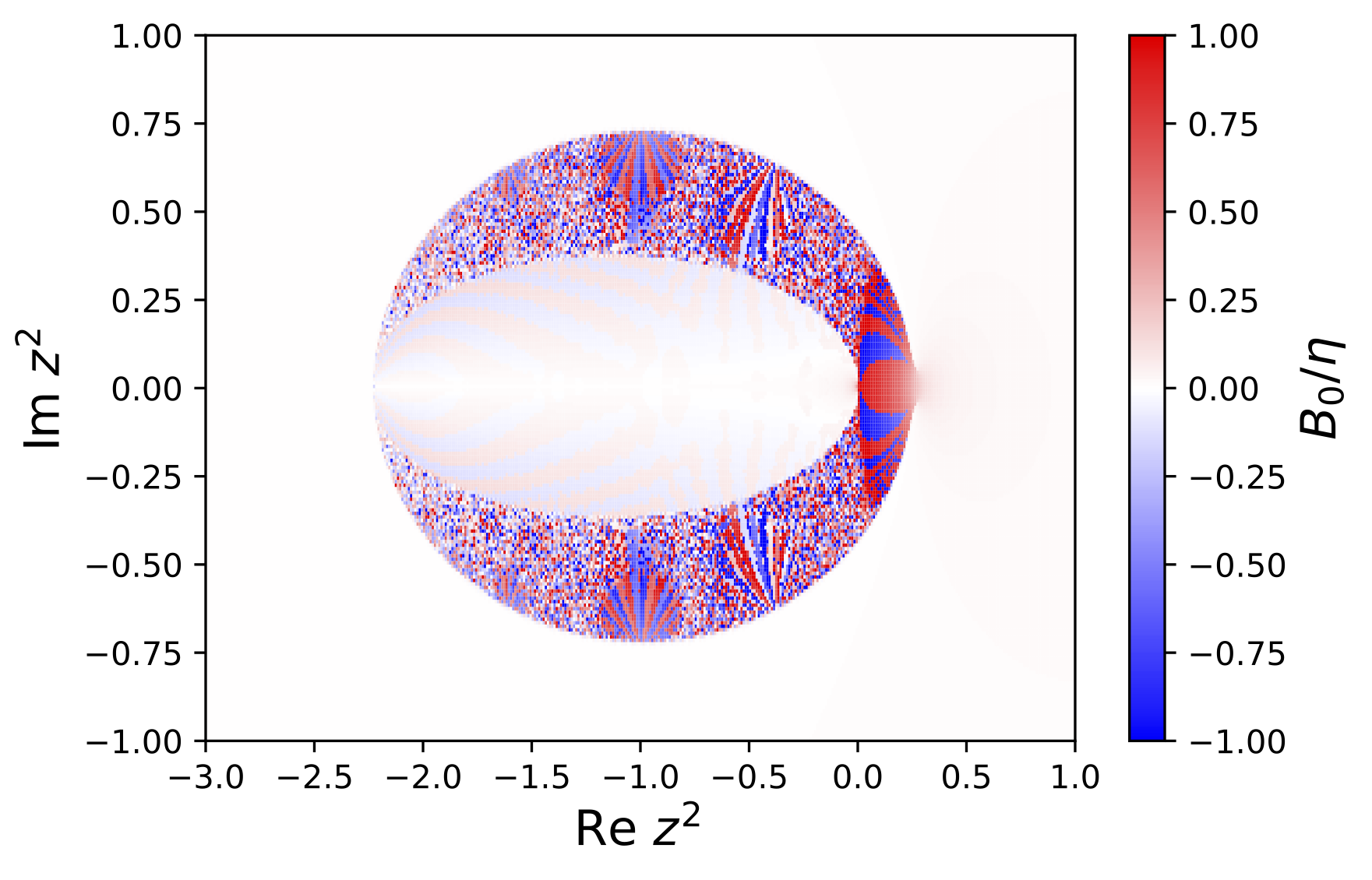}
\includegraphics[width=.5\textwidth]{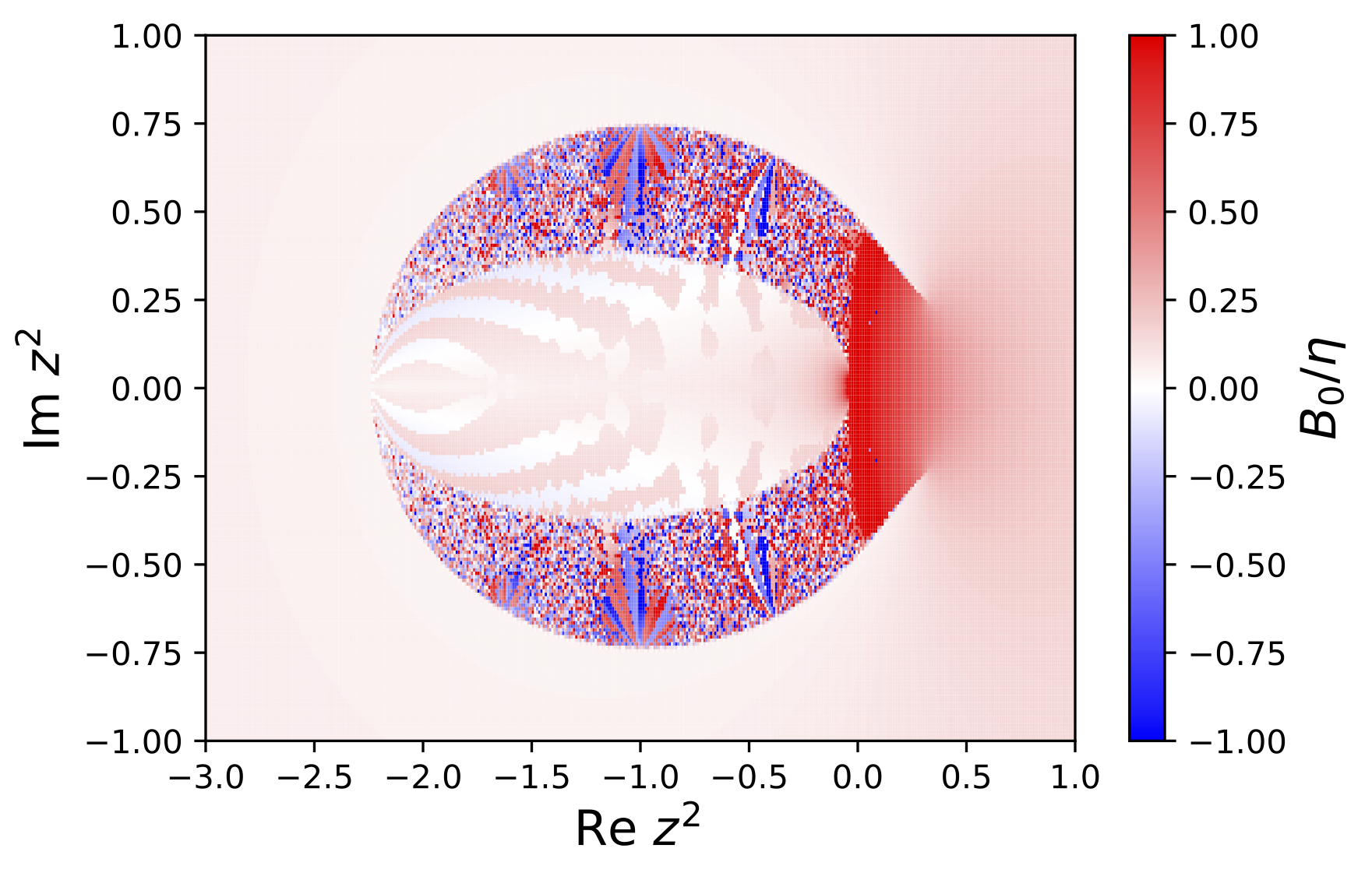}
\caption{\label{fig:rb010_iterated} Real part of the scalar gap $B$ after 300 iterations starting from $A_0=1$ (top and bottom) and $B_0=m=$ (10 MeV (top), 100 MeV (bottom)).}
\end{figure}

We start our iteration from the non-interacting solution ($B_0=m$, $A_0=1$) and treat $\tilde p^2=z_R^2+iz_I^2$ as one would consider the constant $c$ for the Mandelbrot set $z\xleftarrow{} z^2+c$.
For the moment, this reduces the number of independent variables from three ($\vec p^2, p_4, \mu$) to two.
The result of such an iteration is shown in Fig.\ref{fig:rb010_iterated} for the real part of the scalar gap $B$ at two different bare-quark masses of 10 MeV and 100 MeV, respectively.  

\begin{figure}[htb]
\includegraphics[width=.5\textwidth]{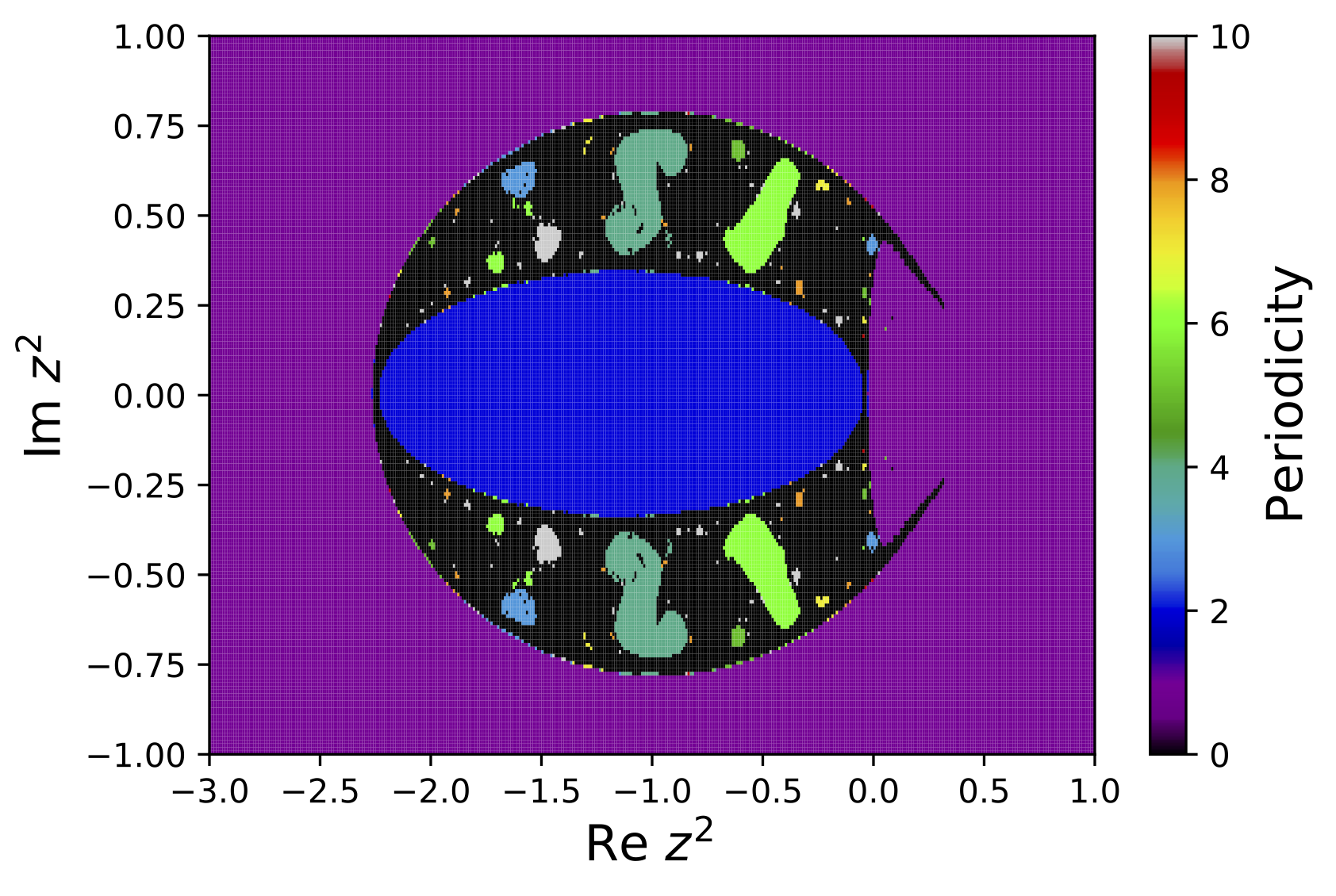}
\caption{\label{fig:period} Periodicity of the iterative mass gap solution at $m=100$ MeV. The outer, indigo region of the plot is absolutely stable under iteration, the inner almond shape has periodicity two, the area in between exhibits chaos with increasing periodicity. For this plot, areas with periodicity larger than ten are plotted in black.}
\end{figure}
\begin{figure}[htb]
\includegraphics[width=.5\textwidth]{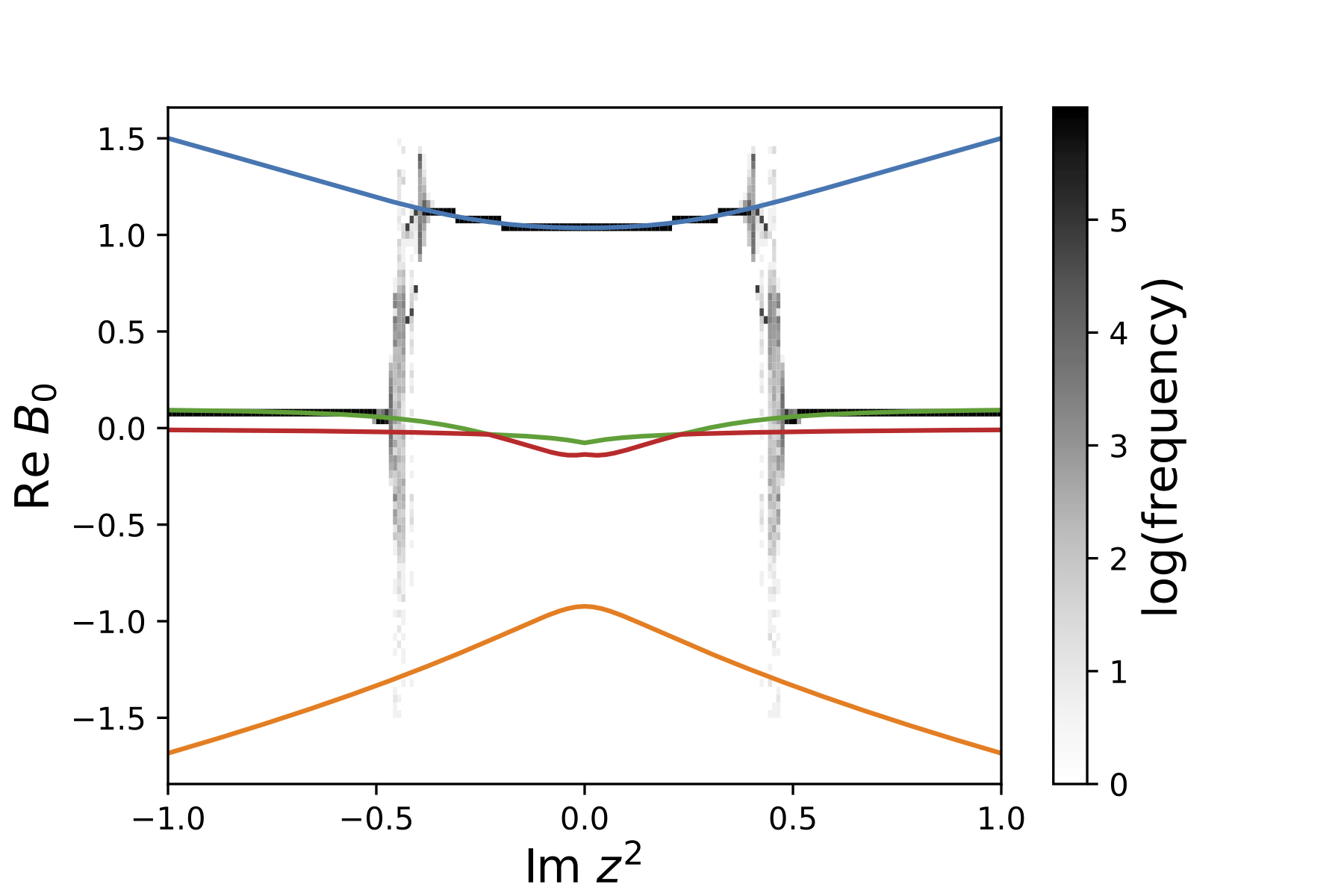}
\caption{\label{fig:zrconst} Real part of the mass gap $B$ at $z_R^2=0.1$ MeV$^2$. The color coding indicates how frequently a solution has been found over 300 iterations after the first 100 iterations which are sufficient to shape the fractal as seen. For reference, all analytic solutions to the polynomial gap equations are plotted in color. Iteration switches from massive solutions (blue) at small $\mathcal I(z^2)$ to bare-mass solutions (green) at larger values. Except for the chaotic transition domain, the iterative approach picks positive mass-gap solutions, only. Note, that the chaotic domain has solutions of periodicity two and higher; it is truly unstable and hence we add a gray scale to measure the frequency of a particular solution over the final 300 iterations.}
\end{figure}

Unlike the Mandelbrot fractal, this fractal does not diverge; chaos exhibits in domains in which the gaps for infinitesimally changes of energy and momentum take vastly different but finite values in a seemingly random pattern. This fractal region is contained within an almost perfectly shaped ellipsoid which we fit accordingly with
\begin{equation}
    {\left( \frac{z_{R}^2+z_{R,0}^2}{R_{R}^2} \right)}^2+{\left( \frac{z_{I}^2}{R_{I}^2} \right)}^2=1.
\label{eqn:chaosellipse}
\end{equation}
$(z_{R,0}^2,R_{R}^2,R_{I}^2)$ differs slightly for $m=10$ MeV $(0.98,1.26,0.77)$ MeV$^2$ and for $m=100$ MeV $(0.99,1.28,0.80)$ MeV$^2$.
The inner almond shape with less obvious chaotic behavior is well approximated by the same function with  $(1.115,1.085,0.310)$ MeV$^2$ for $m=10$ MeV, and 
$(1.150,1.100,0.340)$ MeV$^2$ for $m=100$ MeV.
As for the Mandelbrot set one would be ill advised to understand these figures as a valid representation of the fractal; the appearance of the fractal changes with each new iteration.
We identify regions with identical periodicity, ranging from a stable, period one solution in the region outside of the covering ellipse over a period two region within the almond shape up to higher and higher periodicity in between these two regions.
This is illustrated in Fig.\ref{fig:zrconst} for  $m=100$ MeV for a periodicity of up to ten.

Keeping in mind, that the analytic gap equations possess four distinct solutions it seems interesting, that there is an extended stable domain (periodicity one) which favors one, and only one solution.
We follow the gap solution along a vertical path at fixed $z_R^2$ and vary $z_I^2$.
For reasons which become more clear at later stage we chose $z_R^2=0.1$ MeV$^2$. Along this line, one notices that one passes from an outer stable region into an inner stable region by traversing a small chaotic domain. This is illustrated in Fig.\ref{fig:zrconst}. As within this chaotic domain the value of the gap function can change with each iteration, we plot all obtained values of $\mathcal R (B)$  over 300 iterations in gray scale according to how frequently a particular solution has been obtained.
Evidently, there is a transition between two distinct analytic solutions of the non-iterated forth order polynomial gap equations.
This result seems remarkable if one recalls how one would usually deal with different gap solutions for a given model: each solution is understood as a distinct phase, then one examines the stability of each individual solution and picks the energetically favored solution as the physical one.
Upon iteration we are lead to a different conclusion. Although each of the analytic solution indeed is a solution of the gap equations, only one of them can be stable upon iteration at a given energy and momentum. However, the stable iterative solution over a finite range of energies can switch between distinct analytic solutions. It is further remarkable, that the iteratively stable solution is massive at low and massless at high energy. Amongst all the possibilities chaos seems to offer, this seems a very reasonable one.
The notion of analytic solutions describing different phases however, is not supported from an iterative perspective; there is one, and only one, iterative solution to the gap equation.

\begin{figure*}[htbp]
        \centering
        \includegraphics[width=\textwidth]{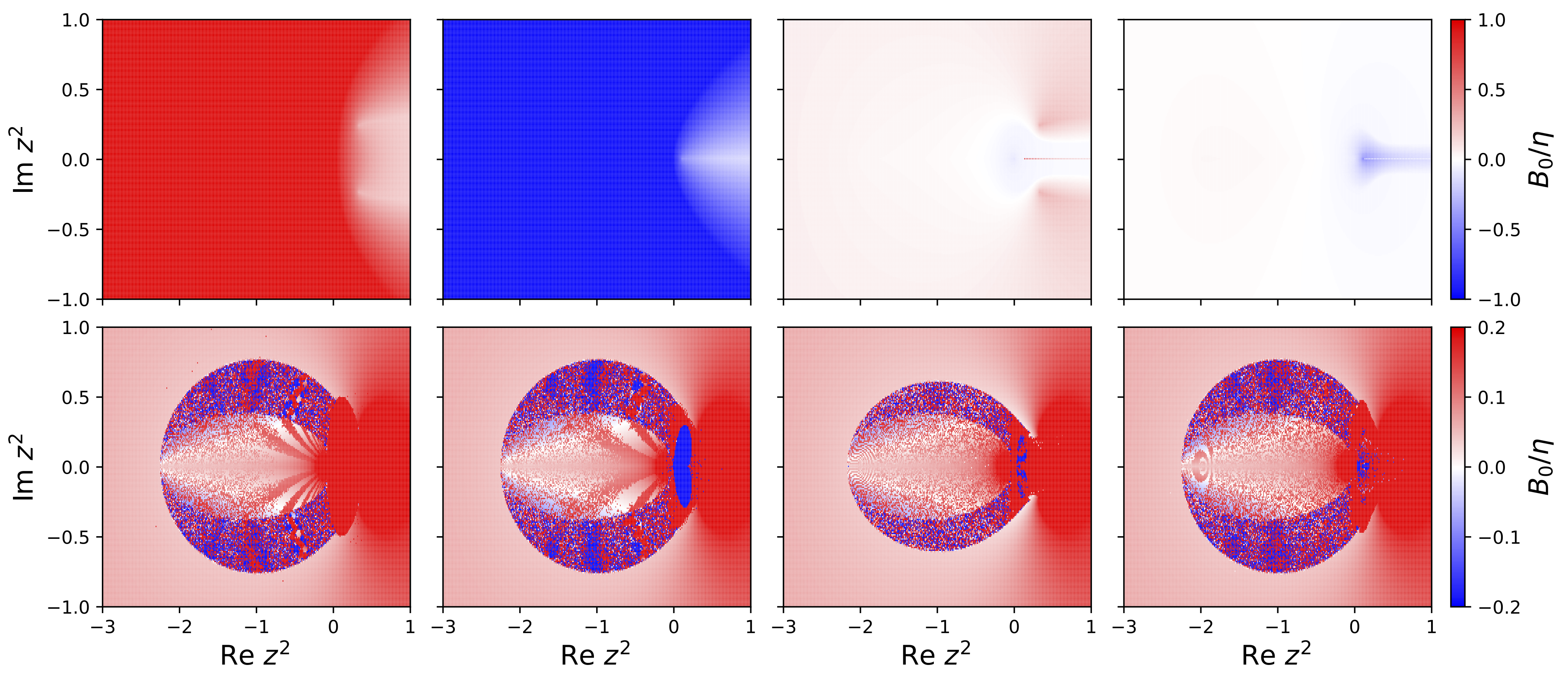}
     \caption{Upper panel: Solutions of the polynomial gap equations for $m=100$ MeV . Each is plotted on a scale that most accentuates its structure. Solution 1 and 3 (from the left) are stable in some, mutually exclusive domains under iteration as illustrated in Fig.\ref{fig:zrconst}. Lower panel: After 300 iterations using the corresponding solution of the polynomial gap equations from the upper panel as initial seed for the iteration. In the outer, non-chaotic domain all four cases produce nearly identical results with positive mass gap only.}
    \label{fig:analyticm100}
\end{figure*}
Before we go into further interpretation of what this result implies, we wish to address a question related to the previous paragraph. Initially we remarked that our iteration starts from the non-interacting solution $A=1, B=m$.
As we try to proceed as careful as possible, let us investigate the iterative stability of the four analytic gap solutions as plotted in the upper panels of Fig.\ref{fig:analyticm100} where we show again the real part of the mass gap.
The lower panel of Fig.\ref{fig:analyticm100} shows the result after 300 iterations of these algebraic solutions as initial value. It is safe to say, that none of them is stable under iteration.
Further, there is a visibly favored solution at large values of $z_R^2$ which does not depend on the initial gap that seeded the iteration.
From a global perspective, the fractal keeps the general shape but shows differences for each different seed solution. This is to be expected and would happen in similar fashion to the Mandelbrot set if the initial value is arbitrarily changed.

\begin{figure*}[phtb]
\includegraphics[width=\textwidth]{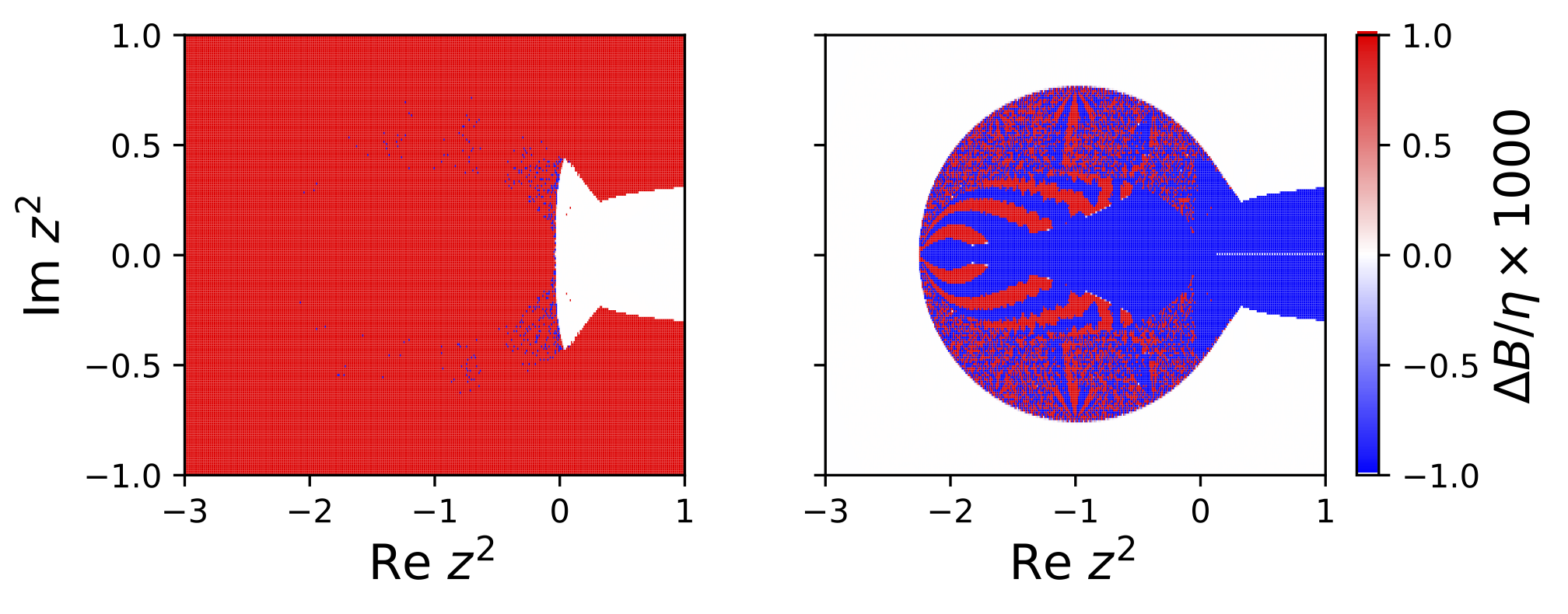}  
\caption{\label{fig:DiffItAn} Difference between gap solution 1 and 3 (from the left)in the top panel of Fig.\ref{fig:analyticm100} and iterative solutions seeded with the non-interacting solution ($A=1$, $B=m$) after 500 iterations.
White domains show no difference between iterative solutions seeded with an analytical model solution or seeded with the non-interacting solution.
Solution 2 and 4 show no agreement anywhere in the stable domain of periodicity one (not shown).}
\end{figure*}

Comparing the iterations to the algebraic solutions of the gap equations in the upper panel of Fig.\ref{fig:analyticm100} one can graphically identify which of them is stable under iteration and in which domain. As seen, this is the case only for the positive mass-gap solutions 1 and 3 from Fig.\ref{fig:analyticm100}, as illustrated in Fig.\ref{fig:DiffItAn}.
In other words, although the chaotic domain will vary, the described features of Fig.~\ref{fig:zrconst} with respect to the analytic gap solutions do not critically depend on the chosen initial gap.

\section{Mass Poles}
\label{SEC:POLES}

\begin{figure*}[phtb]
        \centering
      \includegraphics[width=\linewidth]{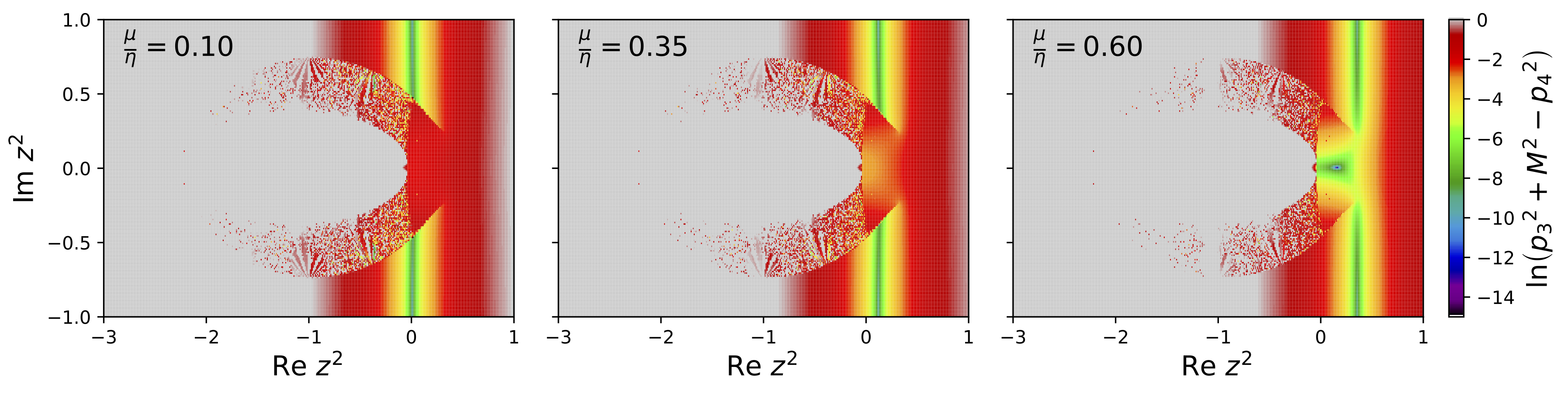}
   \caption{Natural logarithm of $\left(p_3^2+M^2-p_4^2\right)^2$ for the iterative solution for $\mu=(100,350,600)$ MeV (top down) at quark-bare mass $m=$ 100 MeV. The vertical line shaped by minimal negative values indicate a physical mass pole, viz. a quasi-particle. In the chaotic domain, this pole structure is absent, viz. the vertical line (or any distinct pole) pattern is absent. This implies an infrared energy gap below which quarks show no quasi-particle properties. As the chemical potential increases, the quasi-particle pole line moves to the right and simultaneously decreases the gap, viz. the gap region without a pole traces the outer shape of the fractal. Once the chemical potential is sufficiently large, the gap closes entirely. Note, that the absence of a mass pole does not imply that there is no mass gap solution, as illustrated in Fig.\ref{fig:zrconst}.}
    \label{fig:poles}
\end{figure*}

Up to this point we refrained from searching for meaning in our study. In spite of the fact that iterative mass gap solutions result in a large domain of chaotic behavior, which may or may not hide future surprises, we cannot help but wonder whether the switching between massive and mass-less gap solutions in the stable domains offers meaning.
Before we go further, we want to recall that MN is considered to be a confining model. This is seen by the fact that the inverse propagator has no roots in the chirally broken phase and therefore integration over four-momentum does not pick up weight to generate a finite particle number. Hence, although confined quarks generate mass via chiral symmetry breaking, the absence of a mass pole results in the absence of a dispersion, viz. there is no explicit relation between specific momenta and energy.
For the vacuum MN model, this is easily understood by the realization that in Minkowski metric
$p^2+M^2(p^2)$ has no real root if at any $p^2$, $M^2>-p^2=p_4^2-\vec p^2$.
This running away of the mass in the chirally broken phase is exactly what happens in the MN model.
However, as we have shown in the previous section iteration erases the distinction between chirally broken and restored phases and suggests that instead there might be a discontinuous gap solution which is  confinement-like affected by dynamical chiral symmetry at small momenta, and at large momenta  chirally unconfined-like and chirally restored. The transition between these domains is characterized by chaotic and unstable solutions (see Fig.~\ref{fig:zrconst}).

At finite chemical potential the real poles of the propagator 
$A^2(\vec p^2-(p_4+i\mu)^2)+B^2$ are 
represented by $\vec p^2 - p_4^2+ \mu^2  + \Re(M^2) = 0$, with $M=B/A$.
We note that the shift of the pole due to chemical potential should not be confused with the physical mass pole of the particle. 
This becomes evident if one considers an ideal non-interacting gas with $M$ being constant and real valued. For the purpose of this study we refer to the physical mass pole, defined by
$\vec p^2 - p_4^2 + \Re(M^2) = 0$.
From the definition $\tilde p^2 = z_R^2+i z_I^2$ we identify
the pole position in our contour plots as $z_R^2=\mu^2$ and $z_I^2=-2 p_4\mu$
for an ideal particle with constant and real $M$. This represents a vertical line in our plots which does not depend on momentum and measures energy with increasing distance from the real axis.
It shifts to higher $z_R^2$ with increasing chemical potential.

In Fig.\ref{fig:poles} we trace the physical mass pole in Minkowski metric by plotting the logarithm of the quantity $\left(p_3^2+M^2-p_4^2\right)^2$ which gives zero and hence a large negative logarithm at the physical mass pole. As the vertical axis does not depend on the mass ($z_I^2=2p_4\mu$) a vertical pole line indicates constant dressed quark masses. 
We observe the absence of such a well ordered pole structure within the chaotic domain. Since the vertical axis is a measure of the particle energy at fixed chemical potential one can conclude that the transition to the massive solution (Fig.~\ref{fig:zrconst}) suppresses quasi-particle behavior in the infrared domain of the model. 
Again, we can trace the physical pole indicated by the vertical line and find $z_R^2=\mu^2-M^2$ since the pole is found at $p_4^2=p_3^2+M^2$.
Following our elliptic fit of the outer boundary of the fractal domain this allows to determine the critical chemical potential where the infrared energy gap entirely disappears
\begin{equation}
    \mu_{C,IR}=\sqrt{m^2+R_R^2-z_{R,0}^2}\quad.
\end{equation}
We find $\mu_{C,IR}\approx 625 MeV$ for $m=100 MeV$ and $\mu_{C,IR}\approx 540 MeV$ for $m=10 MeV$. At these chemical potentials and beyond, quarks can be considered as completely chirally restored.

In order to estimate when mass-pole states can be occupied, we determine at which chemical potential the energy $p_4$ and the Fermi energy or chemical potential $\mu$ turn equal, that is when $z_I^2=\mu^2$ on the elliptic boundary of the fractal at the position of the physical mass-pole with $M=m$. We choose this scenario as this is the critical potential starting from where the particle energy is larger than the chemical potential and thus large enough to populate quasi-particle states. 
This is the case when
\begin{equation}
    \left(\frac{\mu^2-m^2+z_{R,0}^2}{R_R^2}\right)^2+\left(\frac{\mu^2}{R_I^2}\right)^2=1 \quad,
\end{equation}
and holds for the light quark with $m=10 MeV$ at $\mu_m\approx 359 MeV$, for the heavy quark with $m=100 MeV$ at $\mu_m\approx 432 MeV$.

Although this is not a rigorous statement, one can roughly relate the critical chemical potential for the transition from a chirally broken into the restored phase to the in-vacuum dressed-quark mass. 
In our case, the situation is a bit different. We estimate a hypothetical chirally broken quark vacuum mass based on the previous estimate of the critical potential for the complete disappearance of the infrared gap by setting them approximately equal. 
Relating $\mu_m$ as onset of a deconfined, chirally restored quark phase with an estimate of the constituent quark mass seems to give rather reasonable results in comparison to other model calculations. This is interesting, considering that in the MN model the vacuum mass at zero 4-momentum is defined by the coupling strength $\eta$, which is of the order of 1 GeV.

It is noteworthy that our simple approach reproduces quantities related to the effective constituent masses at reasonable values. We state explicitly, that in this model constituent masses are nowhere realized for a physical particle, viz. an entity with a mass pole of that magnitude.
We can compare the light quark critical chemical potential $\mu_m\approx 359 MeV$ with the deconfinement critical potential obtained within the MN model in Euclidean metric with a value of $300$ MeV \cite{Klahn:2009mb} or with subsequent work based on a widened version of the effective gluon propagator \cite{Chen:2008zr} which predicts deconfinement at a chemical potential of $380 MeV$. There is a satisfying agreement of these values with ours. We point out though, that both of these models are defined within a different metric, slightly different bare quark masses and, most importantly, are based on entirely different assumptions. While the two previous papers employed distinct gap solutions and compare the pressure of the corresponding mass-less Wigner and massive Nambu phase, our approach results in only one gap solution which exhibits a transition from the Nambu to the Wigner phase through a chaotic domain as depicted in Fig.~\ref{fig:zrconst}.
Our quarks are either bare-mass quarks with poles or entities with a chaotic mass function or a dressed quark mass different from the bare mass with no associated pole. In the latter case there is a chaotic transition from dressed quark masses to bare quark masses with increasing energy.

\section{Finite Interaction Width}
\label{SEC:WIDTH}
\begin{figure*}[htbp]
        \includegraphics[width=\linewidth]{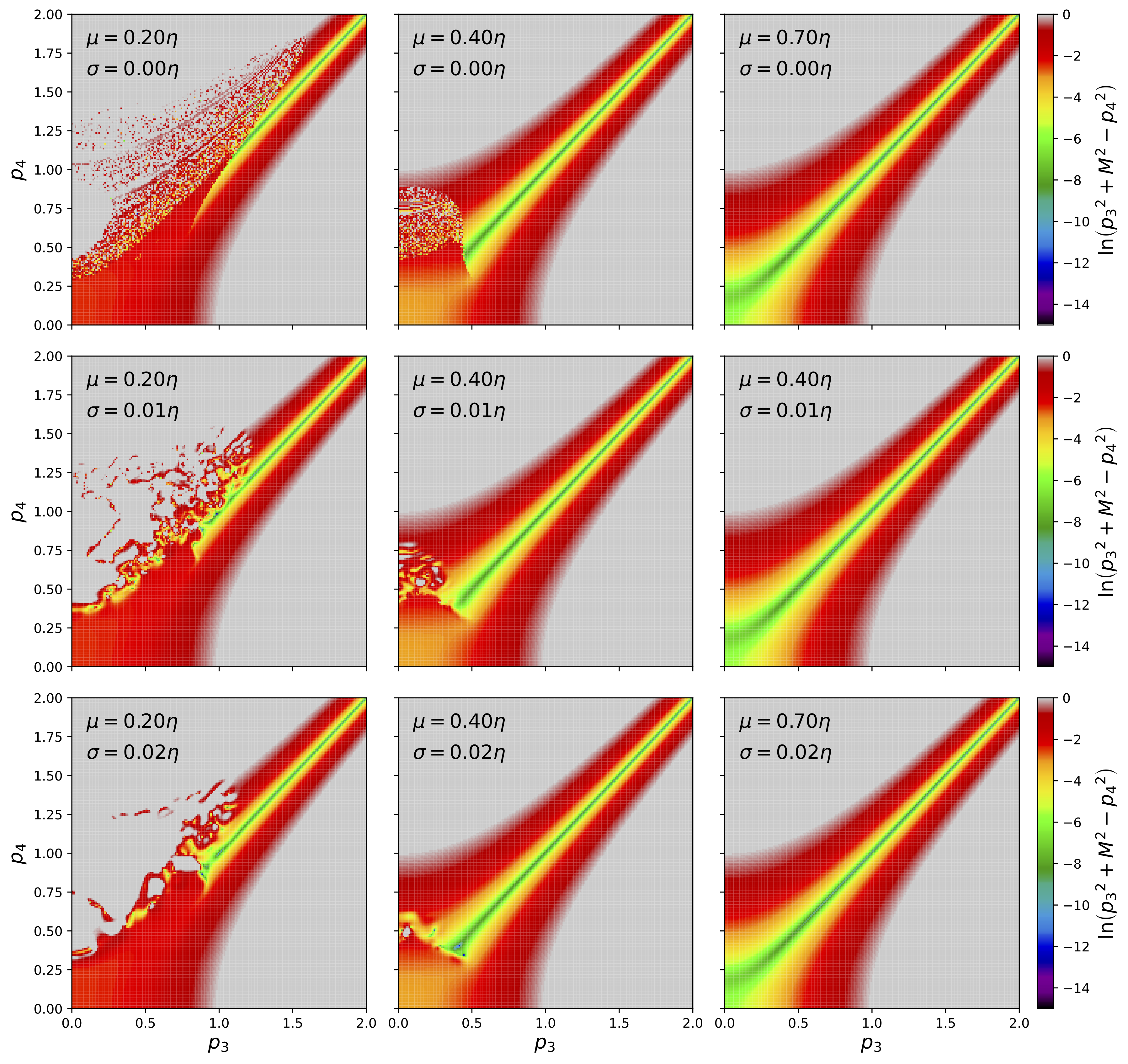}
    \caption{Plotted is the logarithm of the mass-pole condition $\log(|\vec p^2 - p_4^2+ \mu^2  + \Re(M^2)|)$ which shows a dispersion relation with distinct, chaos-induced infrared cut-off. With increasing chemical potential ($m=0.1\eta$; $\mu=(0.2, 0.4, 0.7)\eta$  from left to right) the infrared cut-off decreases and eventually disappears. With increasing widening ($\sigma=(0.00, 0.01, 0.02)\eta$  from top to bottom) chaotic domains blur but the observed IR cut-off remains.} 
    \label{fig:poleplotsp3p4}
\end{figure*}

We begin the final section of this paper with a plot of the particle pole in energy momentum space which we obtain by transforming $(z_R^2,z_I^2)$ to ($p_3=|\vec p |,p_4$) coordinates under explicit choices of the chemical potential as noted in Fig.\ref{fig:poleplotsp3p4}.
Although this switch in representation does not provide additional information we find it instructive to provide an actual dispersion relation obtained from the iterative approach.
In this example at a chemical potential of 700 MeV no chaotic behavior is visible and the dispersion is exactly that of a free quark at bare-mass 100 MeV.
With decreasing chemical potential chaos emerges at energies higher than that of the expected (now absent) free particle dispersion. The actual dispersion branch is cut clean at some critical value (as we discussed in the previous section), thus illustrating our interpretation of the fractal boundary as the cause for a dynamical infrared cutoff below which quarks are mass-pole free.

We address a last question which relates to the fact that the MN model bases on the very particular choice of the effective gluon propagator as a $\delta$-function in 4-momentum space.
The reason that we could easily perform the presented study bases on this Ansatz and the subsequent decoupling of momenta which allows to iterate point-wise for any given 4-momentum without coupling to other momenta.
This might raise the suspicion, that momentum coupling could destroy the fractal structure we observed. In order to keep the simplicity of the gap equations but still get an idea about the stability of the emergent fractal we averaged at each point in our plane after each iteration step and thus mimicked some kind of momentum coupling.
The averaging is based on Gaussian weights around a given point according to
\begin{equation}
    g^2 D^{\mu\nu}(k)=3\pi^4 \eta^2 \delta^{\mu\nu}\frac{\exp(k^2/w^2)} {\int \exp(k^2/w^2)d^4k},
\label{widegluonprop}
\end{equation}
where $w$ is the width of the Gaussian. 
To further simplify, we assumed that the widening only happens in the direction of momentum and energy, i.e. there is no widening perpendicular to the momentum. 

As seen in Fig.\ref{fig:poleplotsp3p4} the separation into chaotic pole-free, and non-chaotic mass-pole domains remains, even when we change the momentum dependence of the gluon from a delta function to a gaussian with a half width as much as 0.02 times the gluon mass. We find numerical evidence that this feature remains even with a width as much as 20\% of the gluon mass, $\approx$ 200 MeV. This corresponds to a spatial width of about 1 fm; about the size of a proton.
Based on this - certainly simplified - treatment of momentum coupling we conclude, that the statements we make in this paper may indeed survive a more complete treatment, involving self-interactions with globally coupled momenta - which has been our main concern prompting this final analysis.

\section{Conclusions}
\label{SEC:CONCL}
As we have shown, a strictly iterative solution of the MN gap equations results in fractal gap structures which can be characterized by the existence of three qualitatively very different, yet co-existing domains of a single and unique gap solution: a bare-mass quark quasi-particle domain with physical mass-poles extending infinitely into the ultraviolet, a dressed-mass quark domain without mass-poles and hence no quasi-particle interpretation in the infrared, and a chaotic domain of transition between the first two phases. Remarkably, the two non-chaotic domains correspond to distinct analytic solutions which would usually represent individual phases with either dynamically broken or restored chiral symmetry.
The fractal approach offers an alternative to this separation which is rooted in the iterative nature of the gap equation.

It is further noteworthy, that the iterative mass gap solution is always positive in the smooth, viz. non-chaotic domain of the fractal.
The appearance of a chaotic boundary between two qualitatively different domains results in interesting properties:\\ 

{\bf I)} The iterative approach provides an ultraviolet cut-off for the massive and mass pole free Nambu solution, as this solution appears only within the elliptic region of the ($z_R^2,z_I^2)$) plane. Thus the approach avoids the appearance of an infinitely increasing dressed-quark mass with increasing momentum and energy. In the MN model, this running mass results in the absence of mass poles for the massive gap solution and thus relates to confinement.\\

{\bf II)} It provides an infrared cut-off for the bare-quark mass Nambu solution and thus ensures that quasi-particle states are not populated at small chemical potential although the quark can {\it virtually} exist as a quasi-particle with well defined dispersion. \\

{\bf III)} Both cutoffs more or less coincide (as seen in Fig.\ref{fig:DiffItAn}) although there is a transition region which is chaotic in nature. The resulting effective Nambu-UV/Wigner-IR cutoff depends dynamically on energy, momentum, bare mass, and chemical potential. As a side note we add that plotting gap solutions in the ($z_R^2,z_I^2$) plane removes much of the dynamical arbitrariness and leaves the ratio of bare mass $m$ and coupling constant $\eta$ as the only 'true' degree of freedom, viz. a change of the chemical potential $\mu$ would rescale the plot but cause no qualitative change, whereas plots like Fig.{\ref{fig:rb010_iterated} show indeed 'the' gap solution at arbitrary chemical potential.\\

{\bf IV)} At sufficiently large chemical potential bare-quark mass-pole states will form at energies which can be populated and thus physical quarks can exist as quasi-particle excitations. \\

A mechanism with these properties can be interpreted as a deconfinement mechanism.
The appearance of one, and only one, iterative solution of the MN gap equations bears a certain elegance. First, by the very fact that there is only one gap solution with expected properties as the existence of only a positive mass gap, asymptotically restored chiral symmetry and the absence or appearance of physical mass poles. Next, it builds on distinct solutions which one would obtain in the non-iterative approach but gives them a new meaning by slicing them into a single new solution with the aforementioned properties.

A simple treatment of a widened, $\delta$-like gluon-interaction indicates that momentum coupling blurs chaotic domains but does not necessarily change the qualitative results we describe if the widening is moderate.
As this study is an exploration and qualitative in nature we look forward to further analyses of this perspective on understanding confinement and the deconfinement transition as highly non-linear and to a certain extent possibly chaotic phenomena.

\section{acknowledgements}
We are grateful to Prashanth Jaikumar, Pok Man Lo, and Craig D. Roberts for helpful comments and discussions which helped us a great deal to find order in chaos.
\bibliography{qcdchaos}

\providecommand{\noopsort}[1]{}\providecommand{\singleletter}[1]{#1}%
\begin{thebibliography}{17}%
\makeatletter
\providecommand \@ifxundefined [1]{%
 \@ifx{#1\undefined}
}%
\providecommand \@ifnum [1]{%
 \ifnum #1\expandafter \@firstoftwo
 \else \expandafter \@secondoftwo
 \fi
}%
\providecommand \@ifx [1]{%
 \ifx #1\expandafter \@firstoftwo
 \else \expandafter \@secondoftwo
 \fi
}%
\providecommand \natexlab [1]{#1}%
\providecommand \enquote  [1]{``#1''}%
\providecommand \bibnamefont  [1]{#1}%
\providecommand \bibfnamefont [1]{#1}%
\providecommand \citenamefont [1]{#1}%
\providecommand \href@noop [0]{\@secondoftwo}%
\providecommand \href [0]{\begingroup \@sanitize@url \@href}%
\providecommand \@href[1]{\@@startlink{#1}\@@href}%
\providecommand \@@href[1]{\endgroup#1\@@endlink}%
\providecommand \@sanitize@url [0]{\catcode `\\12\catcode `\$12\catcode
  `\&12\catcode `\#12\catcode `\^12\catcode `\_12\catcode `\%12\relax}%
\providecommand \@@startlink[1]{}%
\providecommand \@@endlink[0]{}%
\providecommand \url  [0]{\begingroup\@sanitize@url \@url }%
\providecommand \@url [1]{\endgroup\@href {#1}{\urlprefix }}%
\providecommand \urlprefix  [0]{URL }%
\providecommand \Eprint [0]{\href }%
\providecommand \doibase [0]{https://doi.org/}%
\providecommand \selectlanguage [0]{\@gobble}%
\providecommand \bibinfo  [0]{\@secondoftwo}%
\providecommand \bibfield  [0]{\@secondoftwo}%
\providecommand \translation [1]{[#1]}%
\providecommand \BibitemOpen [0]{}%
\providecommand \bibitemStop [0]{}%
\providecommand \bibitemNoStop [0]{.\EOS\space}%
\providecommand \EOS [0]{\spacefactor3000\relax}%
\providecommand \BibitemShut  [1]{\csname bibitem#1\endcsname}%
\let\auto@bib@innerbib\@empty
\bibitem [{\citenamefont
  {Mandelbrot}()}]{doi:10.1111/j.1749-6632.1980.tb29690.x}%
  \BibitemOpen
  \bibfield  {author} {\bibinfo {author} {\bibfnamefont {B.~B.}\ \bibnamefont
  {Mandelbrot}},\ }\bibfield  {title} {\bibinfo {title} {Fractal aspects of the
  iteration of $z\to\lambda z(1- z)$ for complex $\lambda$ and $z$},\ }\href
  {https://doi.org/10.1111/j.1749-6632.1980.tb29690.x} {\bibfield  {journal}
  {\bibinfo  {journal} {Annals of the New York Academy of Sciences}\ }\textbf
  {\bibinfo {volume} {357}},\ \bibinfo {pages} {249}}\BibitemShut {NoStop}%
\bibitem [{\citenamefont {Kadanoff}(1986)}]{Kadanoff:1986}%
  \BibitemOpen
  \bibfield  {author} {\bibinfo {author} {\bibfnamefont {L.}~\bibnamefont
  {Kadanoff}},\ }\bibfield  {title} {\bibinfo {title} {{Fractals: Where's the
  Physics?}},\ }\href {https://doi.org/10.1063/1.2814878} {\bibfield  {journal}
  {\bibinfo  {journal} {Physics Today}\ }\textbf {\bibinfo {volume} {39}},\
  \bibinfo {pages} {6} (\bibinfo {year} {1986})}\BibitemShut {NoStop}%
\bibitem [{\citenamefont {of~Stories}(2020)}]{WoS}%
  \BibitemOpen
  \bibfield  {author} {\bibinfo {author} {\bibfnamefont {W.}~\bibnamefont
  {of~Stories}},\ }\href
  {https://www.webofstories.com/play/benoit.mandelbrot/8} {\emph {\bibinfo
  {title} {B.Mandelbrot about: Drawing; the ability to think in pictures and
  its continued influence}}} (\bibinfo {year} {(accessed July,
  2020)})\BibitemShut {NoStop}%
\bibitem [{\citenamefont {Hofstadter}(1976)}]{PhysRevB.14.2239}%
  \BibitemOpen
  \bibfield  {author} {\bibinfo {author} {\bibfnamefont {D.~R.}\ \bibnamefont
  {Hofstadter}},\ }\bibfield  {title} {\bibinfo {title} {Energy levels and wave
  functions of bloch electrons in rational and irrational magnetic fields},\
  }\href {https://doi.org/10.1103/PhysRevB.14.2239} {\bibfield  {journal}
  {\bibinfo  {journal} {Phys. Rev. B}\ }\textbf {\bibinfo {volume} {14}},\
  \bibinfo {pages} {2239} (\bibinfo {year} {1976})}\BibitemShut {NoStop}%
\bibitem [{\citenamefont {Kuhl}\ and\ \citenamefont
  {St\"ockmann}(1998)}]{PhysRevLett.80.3232}%
  \BibitemOpen
  \bibfield  {author} {\bibinfo {author} {\bibfnamefont {U.}~\bibnamefont
  {Kuhl}}\ and\ \bibinfo {author} {\bibfnamefont {H.-J.}\ \bibnamefont
  {St\"ockmann}},\ }\bibfield  {title} {\bibinfo {title} {Microwave realization
  of the hofstadter butterfly},\ }\href
  {https://doi.org/10.1103/PhysRevLett.80.3232} {\bibfield  {journal} {\bibinfo
   {journal} {Phys. Rev. Lett.}\ }\textbf {\bibinfo {volume} {80}},\ \bibinfo
  {pages} {3232} (\bibinfo {year} {1998})}\BibitemShut {NoStop}%
\bibitem [{\citenamefont {Roberts}(2016)}]{Roberts:2015lja}%
  \BibitemOpen
  \bibfield  {author} {\bibinfo {author} {\bibfnamefont {C.~D.}\ \bibnamefont
  {Roberts}},\ }\bibfield  {title} {\bibinfo {title} {{Three Lectures on Hadron
  Physics}},\ }\href {https://doi.org/10.1088/1742-6596/706/2/022003}
  {\bibfield  {journal} {\bibinfo  {journal} {J. Phys. Conf. Ser.}\ }\textbf
  {\bibinfo {volume} {706}},\ \bibinfo {pages} {022003} (\bibinfo {year}
  {2016})},\ \Eprint {https://arxiv.org/abs/1509.02925} {arXiv:1509.02925
  [nucl-th]} \BibitemShut {NoStop}%
\bibitem [{\citenamefont {Horn}\ and\ \citenamefont
  {Roberts}(2016)}]{Horn:2016rip}%
  \BibitemOpen
  \bibfield  {author} {\bibinfo {author} {\bibfnamefont {T.}~\bibnamefont
  {Horn}}\ and\ \bibinfo {author} {\bibfnamefont {C.~D.}\ \bibnamefont
  {Roberts}},\ }\bibfield  {title} {\bibinfo {title} {{The pion: an enigma
  within the Standard Model}},\ }\href
  {https://doi.org/10.1088/0954-3899/43/7/073001} {\bibfield  {journal}
  {\bibinfo  {journal} {J. Phys. G}\ }\textbf {\bibinfo {volume} {43}},\
  \bibinfo {pages} {073001} (\bibinfo {year} {2016})},\ \Eprint
  {https://arxiv.org/abs/1602.04016} {arXiv:1602.04016 [nucl-th]} \BibitemShut
  {NoStop}%
\bibitem [{\citenamefont {Eichmann}\ \emph {et~al.}(2016)\citenamefont
  {Eichmann}, \citenamefont {Sanchis-Alepuz}, \citenamefont {Williams},
  \citenamefont {Alkofer},\ and\ \citenamefont {Fischer}}]{Eichmann:2016yit}%
  \BibitemOpen
  \bibfield  {author} {\bibinfo {author} {\bibfnamefont {G.}~\bibnamefont
  {Eichmann}}, \bibinfo {author} {\bibfnamefont {H.}~\bibnamefont
  {Sanchis-Alepuz}}, \bibinfo {author} {\bibfnamefont {R.}~\bibnamefont
  {Williams}}, \bibinfo {author} {\bibfnamefont {R.}~\bibnamefont {Alkofer}},\
  and\ \bibinfo {author} {\bibfnamefont {C.~S.}\ \bibnamefont {Fischer}},\
  }\bibfield  {title} {\bibinfo {title} {{Baryons as relativistic three-quark
  bound states}},\ }\href {https://doi.org/10.1016/j.ppnp.2016.07.001}
  {\bibfield  {journal} {\bibinfo  {journal} {Prog. Part. Nucl. Phys.}\
  }\textbf {\bibinfo {volume} {91}},\ \bibinfo {pages} {1} (\bibinfo {year}
  {2016})},\ \Eprint {https://arxiv.org/abs/1606.09602} {arXiv:1606.09602
  [hep-ph]} \BibitemShut {NoStop}%
\bibitem [{\citenamefont {Burkert}\ and\ \citenamefont
  {Roberts}(2019)}]{Burkert:2017djo}%
  \BibitemOpen
  \bibfield  {author} {\bibinfo {author} {\bibfnamefont {V.~D.}\ \bibnamefont
  {Burkert}}\ and\ \bibinfo {author} {\bibfnamefont {C.~D.}\ \bibnamefont
  {Roberts}},\ }\bibfield  {title} {\bibinfo {title} {{Colloquium : Roper
  resonance: Toward a solution to the fifty year puzzle}},\ }\href
  {https://doi.org/10.1103/RevModPhys.91.011003} {\bibfield  {journal}
  {\bibinfo  {journal} {Rev. Mod. Phys.}\ }\textbf {\bibinfo {volume} {91}},\
  \bibinfo {pages} {011003} (\bibinfo {year} {2019})},\ \Eprint
  {https://arxiv.org/abs/1710.02549} {arXiv:1710.02549 [nucl-ex]} \BibitemShut
  {NoStop}%
\bibitem [{\citenamefont {Fischer}(2019)}]{Fischer:2018sdj}%
  \BibitemOpen
  \bibfield  {author} {\bibinfo {author} {\bibfnamefont {C.~S.}\ \bibnamefont
  {Fischer}},\ }\bibfield  {title} {\bibinfo {title} {{QCD at finite
  temperature and chemical potential from Dyson\textendash{}Schwinger
  equations}},\ }\href {https://doi.org/10.1016/j.ppnp.2019.01.002} {\bibfield
  {journal} {\bibinfo  {journal} {Prog. Part. Nucl. Phys.}\ }\textbf {\bibinfo
  {volume} {105}},\ \bibinfo {pages} {1} (\bibinfo {year} {2019})},\ \Eprint
  {https://arxiv.org/abs/1810.12938} {arXiv:1810.12938 [hep-ph]} \BibitemShut
  {NoStop}%
\bibitem [{\citenamefont {Roberts}\ and\ \citenamefont
  {Schmidt}(2020)}]{Roberts:2020udq}%
  \BibitemOpen
  \bibfield  {author} {\bibinfo {author} {\bibfnamefont {C.~D.}\ \bibnamefont
  {Roberts}}\ and\ \bibinfo {author} {\bibfnamefont {S.~M.}\ \bibnamefont
  {Schmidt}},\ }\bibfield  {title} {\bibinfo {title} {{Reflections upon the
  Emergence of Hadronic Mass}}\ }(\bibinfo {year} {2020})\ \Eprint
  {https://arxiv.org/abs/2006.08782} {arXiv:2006.08782 [hep-ph]} \BibitemShut
  {NoStop}%
\bibitem [{\citenamefont {Qin}\ and\ \citenamefont
  {Roberts}(2020)}]{Qin:2020rad}%
  \BibitemOpen
  \bibfield  {author} {\bibinfo {author} {\bibfnamefont {S.-x.}\ \bibnamefont
  {Qin}}\ and\ \bibinfo {author} {\bibfnamefont {C.~D.}\ \bibnamefont
  {Roberts}},\ }\bibfield  {title} {\bibinfo {title} {{Impressions of the
  Continuum Bound State Problem in QCD}},\ }\href@noop {} {\  (\bibinfo {year}
  {2020})},\ \Eprint {https://arxiv.org/abs/2008.07629} {arXiv:2008.07629
  [hep-ph]} \BibitemShut {NoStop}%
\bibitem [{\citenamefont {Barabanov}\ \emph {et~al.}(2021)\citenamefont
  {Barabanov} \emph {et~al.}}]{Barabanov:2020jvn}%
  \BibitemOpen
  \bibfield  {author} {\bibinfo {author} {\bibfnamefont {M.}~\bibnamefont
  {Barabanov}} \emph {et~al.},\ }\bibfield  {title} {\bibinfo {title} {{Diquark
  Correlations in Hadron Physics: Origin, Impact and Evidence}},\ }\href
  {https://doi.org/10.1016/j.ppnp.2020.103835} {\bibfield  {journal} {\bibinfo
  {journal} {Prog. Part. Nucl. Phys.}\ }\textbf {\bibinfo {volume} {116}},\
  \bibinfo {pages} {103835} (\bibinfo {year} {2021})},\ \Eprint
  {https://arxiv.org/abs/2008.07630} {arXiv:2008.07630 [hep-ph]} \BibitemShut
  {NoStop}%
\bibitem [{\citenamefont {Martínez}\ and\ \citenamefont
  {Raya}(2019)}]{Martinez:2019ift}%
  \BibitemOpen
  \bibfield  {author} {\bibinfo {author} {\bibfnamefont {A.}~\bibnamefont
  {Martínez}}\ and\ \bibinfo {author} {\bibfnamefont {A.}~\bibnamefont
  {Raya}},\ }\bibfield  {title} {\bibinfo {title} {{Solving the Gap Equation of
  the NJL Model through Iteration: Unexpected Chaos}},\ }\href
  {https://doi.org/10.3390/sym11040492} {\bibfield  {journal} {\bibinfo
  {journal} {Symmetry}\ }\textbf {\bibinfo {volume} {11}},\ \bibinfo {pages}
  {492} (\bibinfo {year} {2019})},\ \Eprint {https://arxiv.org/abs/1904.02732}
  {arXiv:1904.02732 [hep-ph]} \BibitemShut {NoStop}%
\bibitem [{\citenamefont {Munczek}\ and\ \citenamefont
  {Nemirovsky}(1983)}]{PhysRevD.28.181}%
  \BibitemOpen
  \bibfield  {author} {\bibinfo {author} {\bibfnamefont {H.~J.}\ \bibnamefont
  {Munczek}}\ and\ \bibinfo {author} {\bibfnamefont {A.~M.}\ \bibnamefont
  {Nemirovsky}},\ }\bibfield  {title} {\bibinfo {title} {Ground-state
  $q\overline{q}$ mass spectrum in quantum chromodynamics},\ }\href
  {https://doi.org/10.1103/PhysRevD.28.181} {\bibfield  {journal} {\bibinfo
  {journal} {Phys. Rev. D}\ }\textbf {\bibinfo {volume} {28}},\ \bibinfo
  {pages} {181} (\bibinfo {year} {1983})}\BibitemShut {NoStop}%
\bibitem [{\citenamefont {Klahn}\ \emph {et~al.}(2010)\citenamefont {Klahn},
  \citenamefont {Roberts}, \citenamefont {Chang}, \citenamefont {Chen},\ and\
  \citenamefont {Liu}}]{Klahn:2009mb}%
  \BibitemOpen
  \bibfield  {author} {\bibinfo {author} {\bibfnamefont {T.}~\bibnamefont
  {Klahn}}, \bibinfo {author} {\bibfnamefont {C.~D.}\ \bibnamefont {Roberts}},
  \bibinfo {author} {\bibfnamefont {L.}~\bibnamefont {Chang}}, \bibinfo
  {author} {\bibfnamefont {H.}~\bibnamefont {Chen}},\ and\ \bibinfo {author}
  {\bibfnamefont {Y.-X.}\ \bibnamefont {Liu}},\ }\bibfield  {title} {\bibinfo
  {title} {{Cold quarks in medium: an equation of state}},\ }\href
  {https://doi.org/10.1103/PhysRevC.82.035801} {\bibfield  {journal} {\bibinfo
  {journal} {Phys. Rev. C}\ }\textbf {\bibinfo {volume} {82}},\ \bibinfo
  {pages} {035801} (\bibinfo {year} {2010})},\ \Eprint
  {https://arxiv.org/abs/0911.0654} {arXiv:0911.0654 [nucl-th]} \BibitemShut
  {NoStop}%
\bibitem [{\citenamefont {Chen}\ \emph {et~al.}(2008)\citenamefont {Chen},
  \citenamefont {Yuan}, \citenamefont {Chang}, \citenamefont {Liu},
  \citenamefont {Klahn},\ and\ \citenamefont {Roberts}}]{Chen:2008zr}%
  \BibitemOpen
  \bibfield  {author} {\bibinfo {author} {\bibfnamefont {H.}~\bibnamefont
  {Chen}}, \bibinfo {author} {\bibfnamefont {W.}~\bibnamefont {Yuan}}, \bibinfo
  {author} {\bibfnamefont {L.}~\bibnamefont {Chang}}, \bibinfo {author}
  {\bibfnamefont {Y.-X.}\ \bibnamefont {Liu}}, \bibinfo {author} {\bibfnamefont
  {T.}~\bibnamefont {Klahn}},\ and\ \bibinfo {author} {\bibfnamefont {C.~D.}\
  \bibnamefont {Roberts}},\ }\bibfield  {title} {\bibinfo {title} {{Chemical
  potential and the gap equation}},\ }\href
  {https://doi.org/10.1103/PhysRevD.78.116015} {\bibfield  {journal} {\bibinfo
  {journal} {Phys. Rev. D}\ }\textbf {\bibinfo {volume} {78}},\ \bibinfo
  {pages} {116015} (\bibinfo {year} {2008})},\ \Eprint
  {https://arxiv.org/abs/0807.2755} {arXiv:0807.2755 [nucl-th]} \BibitemShut
  {NoStop}%
\end{thebibliography}%

\end{document}